\documentclass[aps,twocolumn]{revtex4}

\usepackage{amsmath,amssymb,psfrag}
\usepackage[pdftex]{graphicx}
\begin{document}

\title{Low temperature dynamics of kinks on Ising interfaces}

\author{Alain Karma} \author{Alexander E.~Lobkovsky}
\affiliation{Department of Physics, Northeastern University, Boston,
  MA}

\begin{abstract}
  The anisotropic motion of an interface driven by its intrinsic
  curvature or by an external field is investigated in the context of
  the kinetic Ising model in both two and three dimensions.  We derive
  in two dimensions (2d) a continuum evolution equation for the
  density of kinks by a time-dependent and nonlocal mapping to the
  asymmetric exclusion process.  Whereas kinks execute random walks
  biased by the external field and pile up vertically on the physical
  2d lattice, then execute hard-core biased random walks on a
  transformed 1d lattice.  Their density obeys a nonlinear diffusion
  equation which can be transformed into the standard expression for
  the interface velocity $v = M[(\gamma + \gamma'')\kappa+H]$, where
  $M$, $\gamma + \gamma''$, and $\kappa$ are the interface mobility,
  stiffness, and curvature, respectively.  In 3d, we obtain the
  velocity of a curved interface near the $\langle 100 \rangle$
  orientation from an analysis of the self-similar evolution of 2d
  shrinking terraces.  We show that this velocity is consistent with
  the one predicted from the 3d tensorial generalization of the law
  for anisotropic curvature-driven motion.  In this generalization,
  both the interface stiffness tensor and the curvature tensor are
  singular at the $\langle 100\rangle$ orientation.  However, their
  product, which determines the interface velocity, is smooth. In
  addition, we illustrate how this kink-based kinetic description
  provides a useful framework for studying more complex situations by
  modeling the effect of immobile dilute impurities.
\end{abstract}
\maketitle

\section{Introduction}
\label{sec:intro}

Many bulk properties of polycrystals are strongly influenced by the
underlying microstructure.  Much effort goes into predicting the
motion of grain boundaries in response to a variety of driving forces.
Depending on the nature of the grains, their boundaries migrate in
response to applied stresses \cite{winning02:_mechanisms} or magnetic
fields \cite{molodov99:_magnet}, internal forces associated with grain
boundary curvature \cite{herring49}, concentration gradients
\cite{hillert83:_digm}, etc.  Successful models of microstructure
evolution must be supplied with the details of the ways in which the
grain boundaries respond to the driving forces.

Based on general conclusions of non-equilibrium statistical mechanics,
one would expect the interface to have a unique mobility, i.e.,\ a
unique response coefficient to disparate driving forces.  This
conclusion was recently called into question by both experiments in
polycrystals \cite{huang01:_solute,molodov98:_true} and simulations of
Ising interfaces \cite{mendelev02:_driving}.  These works observed
drastically different shapes of shrinking grains.  Grains shrinking
under the influence of capillarity alone were roughly circular whereas
the presence of other driving forces resulted in strongly anisotropic
shapes.  This observation was most simply interpreted in terms of
different interfacial mobilities for different driving forces.  A
resolution of this apparent paradox in the Ising model
\cite{lobkovsky04:_grain_shape}, which does not require a non-unique
mobility, rests with identifying the crucial role of anisotropy in the
calculation of the capillary driving force.  This driving force is the
strongly anisotropic interfacial stiffness \cite{herring49}, i.e.,\ 
the sum $\gamma+\gamma''$ of the excess free-energy of the interface
$\gamma$ and its second derivative with respect to inclination, rather
than $\gamma$ itself which is much less anisotropic.  It turns out
that the reduced mobility, i.e.,\ the product of the capillary driving
force and the bare mobility, is roughly isotropic, and therefore the
grain shape is isotropic as well.  Here we shed further light on the
microscopic mechanism for cancellation of the anisotropies of the
interfacial stiffness and the interfacial mobility in both 2d and 3d.

The precise microscopic mechanisms responsible for the migration of
grain boundaries are complex.  However, there is hope that generic
features near equilibrium are shared by a large class of models of
moving interfaces.  It is with this hope in mind we use the kinetic
Ising model (KIM), introduced in Ref.~\cite{glauber63:ising}, as a
proxy for studying grain boundaries. The KIM is defined by a
collection of spins $s_i = \pm 1$ on a lattice, a total energy which
is a function of this collection, and rules for dynamic evolution of
the spins at some temperature $\beta = 1/kT$.  The energy in the
presence of a magnetic field $H$ is
\begin{equation}
  \label{eq:H}
  E = -J\sum_{\langle ij \rangle} s_i s_j - H \sum_i s_i,
\end{equation}
where the sum in the first term in Eq.~(\ref{eq:H}) is over pairs of
nearest neighbors.  Glauber dynamics \cite{glauber63:ising} is one
possible scheme for evolving the collection of spins in such a way as
to obtain correct distributions in equilibrium.  This model is perhaps
the simplest representation of non-equilibrium dynamics of interfaces.
It can be used to explore effects of lattice anisotropy on the motion
of domain walls driven by magnetic field or capillary forces.  In
addition, domain nucleation and late stages of phase separation can be
addressed within the KIM.  With simple modifications the KIM can be
used to study the phenomenology of interface motion in the presence of
mobile or quenched impurities.

Much is known about the equilibrium behavior of Ising interfaces.  For
example, an exact expression for the interfacial free energy has been
derived in 2d on a square lattice
\cite{rottman81:_exact,avron82:_roughening}.  Approximate expressions
for this free energy and critical amplitudes in 3d have also been
derived \cite{kochmanski99:_ising_asympt,yurishchev97:_critic_ising}.
The non-equilibrium behavior of the KIM is more complicated.  Whereas
several approximate analytic results exist for the mobility of a
domain wall in 2d
\cite{spohn93:_interface,barma92:_dynam_ising,rikvold00:_analytic},
little progress has been achieved is 3d.

Here we construct a simple and intuitive kinetic description of low
temperature domain walls in the KIM based on the kink degrees of
freedom.  The density of kinks is shown to obey a non-linear diffusion
equation, which is equivalent to the law of anisotropic interface
motion driven by curvature and/or an external field derived from the
interface free-energy and mobility.  It is important to emphasize
that, in our kink description, we obtain the law of interface motion
directly in the continuum limit \emph{without} these expressions as
input into our calculation.  Hence, our kink-based theory can be
viewed as a direct microscopic derivation of the law of interface
motion in the low temperature limit of the KIM, free of extraneous
assumptions.  Moreover, the kink-based description is useful for
analyzing more complicated situations.  We illustrate this point both
by extending the analysis of anisotropic interface motion to 3d and by
examining impurity effects in 2d.

Section \ref{sec:2d} of this paper is devoted to the 2d KIM while the
following Sec.~\ref{sec:3d} extends our results to 3d.  In section
\ref{sec:2d_lowT}, we review the existing results concerning the KIM
in 2d with the focus on the non-equilibrium response of an interface
to curvature and magnetic field.  In Sec.~\ref{sec:direct}, we
rederive the velocity of a curved Ising domain wall driven by
capillary forces using kinks as the degrees of freedom responsible for
the motion of the interface.  This description is accurate at low
temperatures when the rate of nucleation of kink-antikink pairs is
small.  We obtain the shape of a shrinking Ising grain analytically in
Sec.~\ref{sec:evolution}.  In the following section \ref{sec:impure}
we illustrate the usefulness of the kink description by considering
the influence of impurities on the grain boundary motion.  In
Sec.~\ref{sec:3d}, we study curvature driven motion in 3d near a high
symmetry singular orientation where the interface can be represented
by a collection of terraces composed of kinks.  This allows us to use
the 2d analytic results to calculate the interface velocity and
therefore the mobility tensor near this symmetry direction. Finally,
conclusions are given in section IV.

\section{Two-dimensional Kinetic Ising Model}
\label{sec:2d}

\subsection{Low temperature expansion of the interface free 
  energy and mobility}
\label{sec:2d_lowT}

Let us summarize the analytical results obtained so far for the KIM,
focusing on the expressions that have been derived for the interface
mobility.  In 2d, the exact interfacial free energy is known
\cite{rottman81:_exact,avron82:_roughening}.  For our purposes it
suffices to write down the first two terms in the temperature
expansion (enthalpic and entropic respectively).  When the spins are
arranged on a square (denoted by a $\square$) lattice of unit lattice
spacing, this energy is
\begin{equation}
  \label{eq:gamma2d}
  \gamma^\square_\mathrm{2d}(\phi)  =  2J\, (c + s) +
  \frac{1}{\beta}\, [c \, \log c + s \, \log s - (c + s) \,\log (c + s)],
\end{equation}
where $c = |\cos\phi|$, $s = |\sin\phi|$, and $\phi$ is the
inclination defined as the angle of the interface normal with respect
to the $\langle 10 \rangle$ axis of the underlying lattice.

When spins flip according to non-conserved Glauber dynamics
\cite{glauber63:ising}, the interface moves to minimize the free
energy of the system which consists of the bulk and the interface
contributions.  Spohn \cite{spohn93:_interface} has derived the sharp
interface continuum description of a domain wall in KIM.  It follows
from his derivation that the normal velocity of the interface $v$ is
the product of the mobility $M$ and a driving force.  From the
continuum description it also follows that in the absence of magnetic
field, the driving force is the product of the mean curvature of the
interface $\kappa$ and the interface stiffness. In 2d, the stiffness
is $\gamma + \gamma''$, where $\gamma''$ denotes the second derivative
of $\gamma$ with respect to $\phi$.  Using a Green-Kubo perturbative
formalism, Spohn obtained the interface mobility in the limit of small
temperature and small driving magnetic field.  The same result was
obtained earlier by Barma \cite{barma92:_dynam_ising} using a mapping
of the dynamics of the low-temperature Ising interface to the
one-dimensional exclusion process.  Rikvold and Kolesik obtained
analytical expressions valid for large fields and temperatures
\cite{rikvold00:_analytic}.  The leading term in the temperature
expansion of the mobility diverges like $1/T$
\begin{subequations}
  \label{eq:M2d}
  \begin{eqnarray}
    M^\square_\mathrm{2d}(\phi) &=& \frac{\beta}{2\tau} \frac{|\sin
      2\phi|}{|\cos\phi| + |\sin\phi|}, \\
    M^\triangle_\mathrm{2d}(\phi) &=& 
    \frac{\beta\sqrt{3}}{2\tau}\, \frac{\sin\phi \,(\cos\phi -
      \frac{1}{\sqrt{3}}\sin\phi)}{\cos\phi +
      \frac{1}{\sqrt{3}}\sin\phi},
  \end{eqnarray}
\end{subequations}
where $\square$ refers to a square lattice and $\triangle$ to a
triangular lattice.  In addition, $\tau$ is the intrinsic time scale
of the Glauber dynamics which is the inverse frequency of the
attempted spin flips.  The triangular lattice formula is valid in the
$\phi\in[0, \pi/6]$ domain and can be extended to the other angles via
an appropriate symmetry transformation.

The above expressions for the interface energy and mobility can be
combined to arrive at the continuum (or mean field) low temperature
equation of motion of the interface driven by curvature $\kappa$ and
magnetic field $H \ll 1/\beta$.  The normal velocity of the interface
is
\begin{equation}
  \label{eq:v}
  v_\mathrm{2d}(\phi) = M_\mathrm{2d}[\kappa(\gamma_\mathrm{2d} +
  \gamma''_\mathrm{2d}) + H] = M^*_\mathrm{2d} \, \kappa +
  M_\mathrm{2d} \, H,
\end{equation}
where the reduced mobility on the square lattice is (see
Appendix~\ref{app:derivation} for the triangular lattice result)
\begin{equation}
  \label{eq:Mstar}
  M^*_\mathrm{2d}(\phi) \equiv
  M^\square_\mathrm{2d}(\gamma_\mathrm{2d} + \gamma''_\mathrm{2d}) =
  \frac{1}{\tau \, (|\cos\phi| + |\sin\phi|)^2}.
\end{equation}

Note that the reduced mobility is roughly isotropic whereas the bare
mobility is strongly anisotropic.  In addition, the reduced mobility
does not diverge in the $T \rightarrow 0$ limit whereas the bare
mobility does.  This happens because the contribution of the enthalpic
term $2J(c + s)$ to the interfacial stiffness evaluates to zero.
Therefore, only the entropic term (which is proportional to T)
contributes to the stiffness.  Moreover, the contribution due to the
entropic term diverges at the high symmetry orientations whereas the
bare mobility vanishes at those orientations in such a way that the
product of the two quantities produces a finite non-zero reduced
mobility.  This behavior is responsible for the nearly circular shape
of a shrinking grain on a hexagonal lattice in
Ref.~\cite{mendelev02:_driving}.

\subsection{Direct calculation of $M^*$ via the dynamics of kinks}
\label{sec:direct}

The basic law of interface motion embodied in Eq.~\eqref{eq:v} is
usually derived using a thermodynamic approach where the interface
free-energy and mobility are computed separately.  This approach, even
though general, lacks intuitive appeal.  Furthermore, it is not simply
extended to more complex situations.  It is therefore worthwhile to
develop an alternative method for deriving Eq.~\eqref{eq:Mstar}
directly from a microscopic picture without the need to compute the
interface free-energy and mobility as intermediate steps.  We develop
such a method based on a low temperature description of the interface
in terms of kinks.  This simple microscopic picture, and the results
obtained for the velocity of a shrinking grain in 2d, provide the
basis for the subsequent incorporation of impurities and the
derivation of an expression for the interface velocity in 3d.

\begin{figure}[htbp]
  \centering
  \includegraphics[width=3.3in]{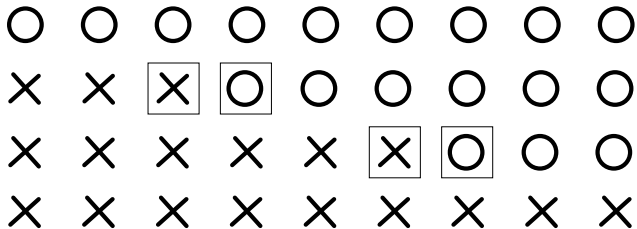}
  \caption{Schematic description of an 
  interface between up (crosses) and down
    (circles) spin domains on a square lattice.  Only corner (boxed)
    spins can flip at low temperature.  A flip of a corner spin
    corresponds to moving the kink left or right.}
  \label{fig:kinks}
\end{figure}

When the temperature is low, $\beta J \gg 1$, the only allowed spin
flips are those that do not increase the total energy.  Therefore
kink-antikink pairs cannot nucleate at the interface.  Barma
\cite{barma92:_dynam_ising} observed that the interface between the
Ising domains can then be represented by a staircase of kinks shown in
Fig.~\ref{fig:kinks}.  The kinetics of kinks reduces to an exclusion
process (asymmetric in the presence of magnetic field)
\cite{liggett85:_interacting,masi84:_nonequilibrium}.  Even though
steady state properties of this process (corresponding to a flat
field-driven interface of a fixed inclination) are well known, little
progress has been made analytically to describe the evolution of a
non-uniform kink distribution corresponding to a curved interface.

\begin{figure}[htbp]
  \centering
  \includegraphics[width=3.3in]{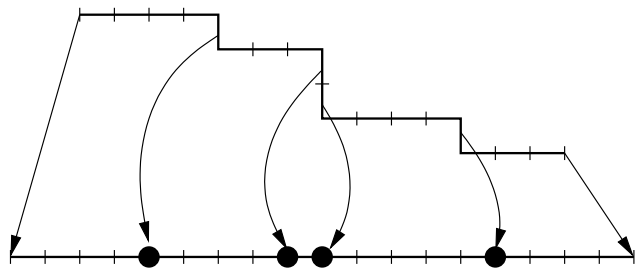}
  \caption{Mapping of the kink dynamics onto the symmetric
  exclusion process corresponding to non-overlapping random walks on a
  1d lattice. Note that kinks that pile up vertically in the
  physical 2d lattice do not overlap in the transformed 1d
  lattice. See Appendix \ref{app:derivation} for details.}
  \label{fig:mapping}
\end{figure}

Let us define the ensemble average density of kinks $\rho(x, t)$ and
derive its evolution equation, which is equivalent to
Eq.~\eqref{eq:v}. We outline the derivation here and relegate the
details to Appendix \ref{app:derivation}.  We focus here on the
curvature driven motion while the appendix includes the effect of the
magnetic field.  Unimpeded by its neighbors, each kink executes a
random walk corresponding to purely diffusive motion.  Many kinks can
``pile-up'' at the same site but cannot pass through each other.  Via
a transformation which inserts an extra lattice site between every
pair of neighboring kinks (illustrated in Fig.~\ref{fig:mapping}), we
map the dynamics of kinks onto the problem of 1d random walkers which
cannot occupy the same lattice site.  The density of walkers for this
symmetric exclusion process obeys a simple diffusion equation
\cite{barma92:_dynam_ising}.  This is true because when two walkers
collide, their indices can be exchanged (i.e. their ``identities''
switched) without affecting their density.  Thus a collision can be
viewed as the tunneling of the kinks through each other without
affecting each other.  The density of hard core random walkers is
insensitive to this identity-switching transformation and must
therefore satisfy a diffusion equation.  When transformed back to the
original coordinate system in which kinks can pile up, the equation
for $\rho(x, t)$ reads for the square lattice (see Appendix
\ref{app:derivation} for the triangular lattice version)
\begin{equation}
  \label{eq:rho_t}
  \tau \rho_t =  
  \left(
    \frac{\rho_x}{(1 + \rho)^2} 
  \right)_x = -F_x = \mu_{xx},
\end{equation}
where subscripts denote differentiation, $F = -\rho_x/(1 + \rho)^2$ is
the flux of kinks and $\mu = -1/(1 + \rho)$ is the kink ``chemical
potential.''  Eq.~\eqref{eq:rho_t} is a nonlinear diffusion equation
with the diffusivity $1/(1 + \rho)^2$ which is a decreasing function
of density.  This reduction results from the fact that when more than
two kinks occupy the same site, some of these kinks are completely
immobile.  Since the density of kinks is defined for interfaces
inclined with respect to the $\langle 10\rangle$ orientation,
Eq.~(\ref{eq:rho_t}) has to be supplemented by boundary conditions
which piece together different $\pi/2$ sectors of the grain boundary.

Geometrically, the density of kinks is the local slope of the
interface with respect to the low energy $\langle 10 \rangle$
orientation.  It is therefore easy to show that Eq.~\eqref{eq:v} with
$H = 0$ and Eq.~\eqref{eq:rho_t} are equivalent (see Appendix
\ref{app:derivation}).  Thus we derived the equation of motion for the
interface without assuming the applicability of the continuum
description of the interface.  Even though, for clarity, we have
restricted our discussion above to motion by curvature only, we derive
in Appendix \ref{app:derivation} the evolution equation for the kink
density for general motion by both curvature and an external field,
and show that it is equivalent to Eq.~\eqref{eq:v}.

Neglecting thermal excitation of kink-antikink pairs allowed us to
construct an equation for a single density of kinks $\rho(x,t)$.  In
general local densities of kinks $\rho_+$ and anti-kinks $\rho_-$ must
be considered.  Each density obeys the non-linear diffusion equation
\eqref{eq:rho_t} augmented by a source term proportional to
$\exp(-2\beta J)$, due to the creation of kink-antikink pairs, and a
sink term proportional to the product $\rho_+\rho_-$ due to the
annihilation of kinks by anti-kinks.  The local slope of the interface
with respect to the low energy orientation is given by the sum $\rho_+
+ \rho_-$ of the kink and anti-kink densities.  Once the details of
this two-density approach are worked out, a formal temperature
perturbation expansion becomes possible since either $\rho_+$ or
$\rho_-$ is exponentially small in low temperature limit.  A small
mobility for the high symmetry orientations, which are immobile at
zeroth order, will be the most important effect at the next order in
the temperature expansion.

\subsection{Evolution of an Ising grain}
\label{sec:evolution}

Let us now use the equation of motion \eqref{eq:v} to describe the
evolution of an ``Ising grain,'' i.e.,\ an island of down spins in a sea
of up spins.  The reduction of the interfacial free energy and spin
alignment parallel to a positive magnetic field are the the driving
forces for the grain shrinkage while negative magnetic field favors
grain growth.  The numerical method of solving \eqref{eq:v} is
described in detail elsewhere \cite{mendelev02:_driving}.

\begin{figure}[htbp]
  \centering
  \includegraphics[width=3.3in]{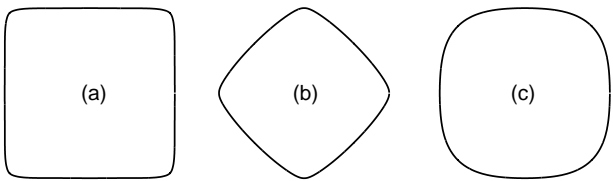}
  \caption{Shape of an evolving Ising domain in 2d.  (a) Final
    stationary shape of a domain in sufficiently strong negative
    magnetic field.  (b) Shape of a shrinking domain in a large
    positive field.  (c) Self-similar shape of a domain shrinking in
    absence of magnetic field.}
  \label{fig:shape}
\end{figure}

When a sufficiently large negative is applied, the grain grows until
it reaches a stationary configuration determined by its initial shape
(see Fig.~\ref{fig:shape}a).  This happens because the velocity of the
interface in the direction of the low energy planes is never outward
since the mobility vanishes for these orientations.  Thus a grain
cannot grow beyond its initial size.  Any growth process has to
include the nucleation of kink-antikink pairs which is explicitly
ignored in our description.

The amplitude of the positive magnetic field $H$ determines the shape
of the shrinking grain.  When $\beta H \gg \beta H_c = 1/R$, the  
second term in Eq.~\eqref{eq:v} dominates.  Note that for large grains
this crossover magnetic field vanishes like $1/R$.  The shape of the
grain shrinking under these conditions, shown in
Fig.~\ref{fig:shape}b, is strongly anisotropic.

When the applied field is much smaller then the crossover field, the
evolution is controlled by the more isotropic reduced mobility and
thus the shape of a shrinking grain is close to a circle.  Even when
initially the dynamics is controlled by the magnetic field, the crossover
to curvature dominated dynamics will happen when the grain shrinks to
a sufficiently small size.  In this regime, the grain shrinks in a
self-similar manner (see Fig.~\ref{fig:shape}c).

\subsubsection{Self-similar evolution of the shrinking grain}

Here we restrict ourselves to the square lattice while citing the
results for the hexagonal lattice in Appendix
\ref{sec:self-similar_hex}.  We also set the time scale $\tau = 1$.

\begin{figure}[htbp]
  \centering
  \includegraphics[width=5cm]{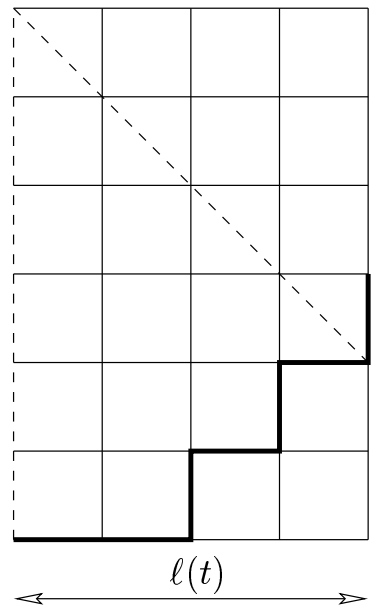}
  \caption{A $\pi/4$ slice of the grain defined by the dashed
    lines that converge in the center of the grain.  The thick line is the
    grain boundary.  The kink density $\rho(x,t)$ is defined on a shrinking
    domain of width $\ell(t)$.}
  \label{fig:pi4slice}
\end{figure}
To compute the shape of a shrinking grain we need to specify the
region in which our kink description holds and fix the boundary
conditions at the edges of this region.  Since we expect the grain to
possess four mirror planes inclined at 0, $\pm \pi/4$ and $\pi/2$ with
respect to the $\langle 10 \rangle$ plane, we will restrict ourselves
to a $\pi/4$ wedge (See Fig.~\ref{fig:pi4slice}).  The slope at the
left edge of the wedge ($x = 0$) is $\rho = 0$ and the slope at the
right edge ($x = \ell(t)$) is $\rho = 1$ due to mirror symmetry around
the $\pi/4$ plane and the smoothness of the grain shape.  Thus we are
to solve Eq.~\eqref{eq:rho_t} subject to the boundary conditions
\begin{equation}
  \label{eq:bc_slice}
  \rho(0, t) = 0, \quad \rho(\ell(t), t) = 1.
\end{equation}

\begin{figure}[htbp]
  \centering
  \includegraphics[width=5cm]{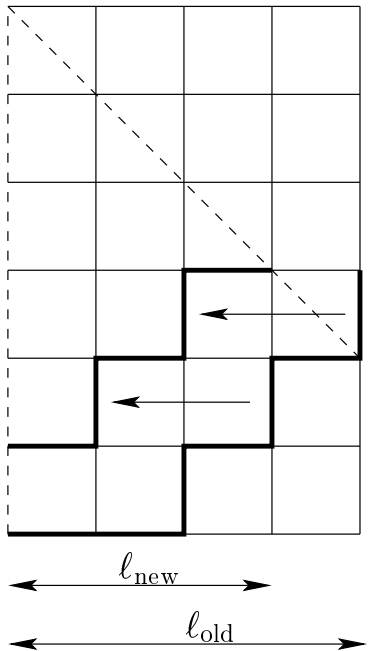}
  \caption{When the rightmost kink moves two steps to the left, the
    width of the domain is reduced by 1.}
  \label{fig:dmain_shrink}
\end{figure}

The final ingredient in determining the grain shape is the shrinking
rate.  The slice width $\ell(t)$ shrinks as the kinks at its right
edge flow to the left with a flux $F(\ell) = -\rho_x(\ell)/4$.  Every
time the kinks move \textit{two} sites to the left, the width of the
slice is reduced by 1 (see Fig.~\ref{fig:dmain_shrink} for a visual
explanation), and therefore
\begin{equation}
  \label{eq:ell_dot}
  \dot \ell(t) = \frac{F(\ell)}{2} = -\frac{\rho_x(\ell)}{8}.
\end{equation}

\begin{figure}[htbp]
  \centering 
  \includegraphics[width=3.3in]{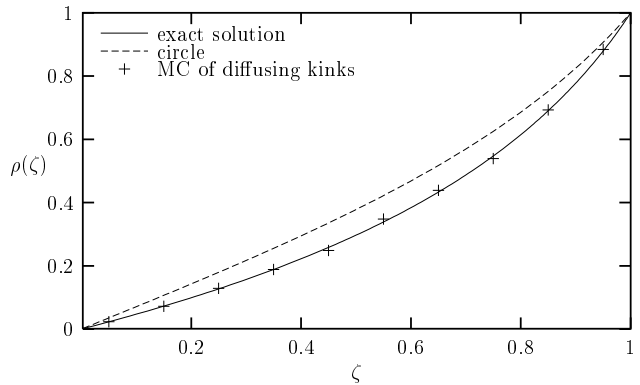}
  \caption{Density of kinks in units of the inverse lattice constant
    plotted against the dimensionless scaled distance $\zeta$ across
    the arc.  The exact density for a self-similar shrinking grain
    obtained by solving \eqref{eq:rho_self_similar} is compared to
    that obtained via a Monte Carlo simulation of diffusing
    impermeable kinks.  The dashed line shows for comparison the
    density of kinks in a circular arc on a square lattice.}
  \label{fig:comparison}
\end{figure}

As we mentioned above, in the absence of a magnetic field, the shape
of the shrinking grain is self-similar.  To prove this we seek a
solution to the moving boundary problem defined by equations
\eqref{eq:rho_t}, \eqref{eq:bc_slice} and \eqref{eq:ell_dot}, which
depends on space and time only through a combination $\zeta =
x/\ell(t)$.  Substituting this Ansatz into the expression for the
shrinking rate \eqref{eq:ell_dot} we obtain
\begin{equation}
  \label{eq:shrinking_rate}
  \ell \dot \ell = -\frac{\rho'(1)}{8},
\end{equation}
where prime denotes differentiation with respect to $\zeta$.  Thus the
the rate of change of the grain area $A \sim \ell^2$ under this
self-similar evolution is constant as expected.  The kink diffusion
equation \eqref{eq:rho_t} becomes
\begin{equation}
  \label{eq:rho_self_similar}
  B \zeta \rho' =
  \left(
    \frac{\rho'}{(1 + \rho)^2}
  \right)',
\end{equation}
with $\rho(0) = 0$, $\rho(1) = 1$, and $B = -\ell\dot \ell =
\rho'(1)/8$.  The constant $B \approx 0.331491$ is determined
self-consistently by a shooting procedure.  Fig.~\ref{fig:comparison}
shows the comparison of the solution of \eqref{eq:rho_self_similar} to
the ensemble averaged Monte Carlo simulation of diffusing hard-core
kinks with boundary conditions appropriate to the shrinking grain
scenario.  Quantitative agreement of the sharp interface result
\eqref{eq:v} with the Monte Carlo simulation of the full KIM was found
in Ref.~\cite{lobkovsky04:_grain_shape}.

It is useful to recast Eq.~\eqref{eq:rho_self_similar} in terms of the
polar parameterization of the self-similar shrinking grain $r(\phi, t)
= \sqrt{2} \, \ell(t) \, r_\mathrm{2d}(\phi)$.  We chose to scale
$r(\phi, t)$ in such a way that $r_\mathrm{2d}(\pi/4) = 1$.  The kink
density of a self-similarly shrinking grain is a function of $\phi$
only
\begin{equation}
  \label{eq:rho_polar}
  \tilde \rho(\phi) = \frac{r_\mathrm{2d}(\phi)\, \sin\phi -
  r'_\mathrm{2d}(\phi) \, \cos\phi}{r_\mathrm{2d}(\phi) \, \cos\phi +
  r'_\mathrm{2d}(\phi) \, \sin\phi}.
\end{equation}
The equation for the shape of the self-similarly shrinking grain in
the polar coordinates can be integrated once to yield
\begin{multline}
  \label{eq:r_2d}
  2B \, r_\mathrm{2d}^2(\phi)
  \left[
    r_\mathrm{2d}(\phi) (\sin\phi + \cos\phi)
    + r'_\mathrm{2d}(\phi) (\sin\phi - \cos\phi)
  \right]^2 \cr
  = r_\mathrm{2d}^2(\phi) + 2 r'_\mathrm{2d}(\phi) - r''_\mathrm{2d}(\phi),
\end{multline}
subject to $r'_\mathrm{2d}(0) = r'_\mathrm{2d}(\pi/4) = 0$ (by
symmetry) and $r_\mathrm{2d}(\pi/4) = 1$.  One of these conditions is
automatically satisfied for the value of $B$ found above.

\begin{figure}[htbp]
  \centering
  \includegraphics[width=3.3in]{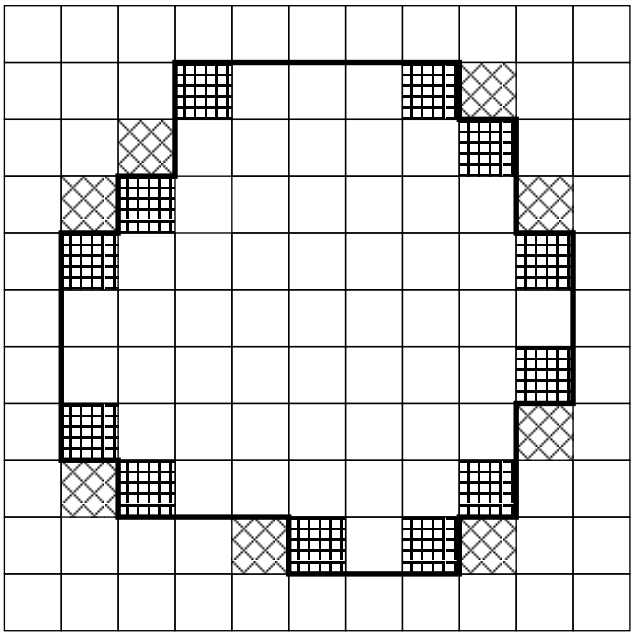}
  \caption{The number of the outside corner spins (diagonal
    crosshatch) for any domain is always smaller by four than the
    number of the inside corner spins (horizontal-vertical
    crosshatch).  The shrinking rate of this domain, computed from the
    difference between the inside and outside corner spins, is thus
    known exactly in the low temperature limit.}
  \label{fig:corner_spins}
\end{figure}

Let us finally mention another analytic result concerning the
grain shrinking rate $dA/dt$
\begin{equation}
  \label{eq:dAdt}
  -\tau \frac{dA}{dt} = \oint d\phi \,
  M^*_\mathrm{2d}(\phi) = 
  \begin{cases}
    4, & \text{square}, \\
    3\sqrt{3}, & \text{hexagonal}.
  \end{cases}
\end{equation}
These formulas (the square lattice result first appeared in
Ref.~\cite{kandel90:_rigorous}) are a simple consequence of the fact
that only the corner spins are allowed to flip (see
Fig.~\ref{fig:corner_spins}).  When a spin in a concave corner flips,
the area of the grain increases by 1 (square lattice).  And vice
versa, when a kink in convex corner flips, the area is reduced by 1.
Since the probabilities of all allowed spin flips are the same and the
number of convex kinks on a square lattice is greater than the number
of concave kinks by 4 (due to Hopf's theorem which states that the
rotation index of a simple curve is 1), we arrive at \eqref{eq:dAdt}.
Ref.~\cite{lobkovsky04:_grain_shape} checked that the shrinking rate
on the hexagonal lattice is indeed $3\sqrt{3} \approx 5.196$.

\subsection{Drag by immobile impurities}
\label{sec:impure}

The kink picture of the low-temperature grain boundary dynamics is
useful in understanding the effect of dilute immobile impurities.  We
model the interaction of the grain boundary with interstitial
impurities by defining a variable $\theta_{mn}$ on the dual lattice
sites.  $\theta = 1$ when an impurity is present and $0$ otherwise.
The impurities are randomly positioned on the dual lattice and do not
move.  The interaction of the impurities with the spins is introduced
via an additional term in the energy
\begin{equation}
  \label{eq:impurities}
  E_\mathrm{imp} = \epsilon \sum_{m,n} \theta_{mn}
  S_{\langle mn \rangle},
\end{equation}
where $S_{\langle mn \rangle}$ is the total magnetization of the Ising
spins nearest to the impurity located at site $(m,n)$ of the dual
lattice and the sum is over all the dual lattice sites.

\begin{figure}[htbp]
  \centering
  \includegraphics[width=3.3in]{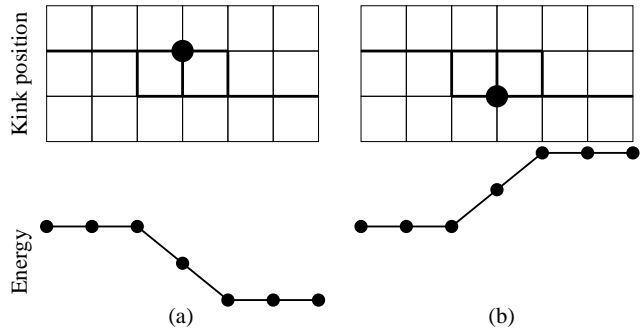}
  \caption{Illustration of the energy landscape resulting from the
    interaction of a kink with a single fixed impurity (large solid
    circle) on the dual lattice. The up (down) spin domain is above
    (below) the interface represented by a thick black line along the
    dual lattice.  The vertical segments denote three positions of the
    kink.  The energy (in arbitrary units) corresponding to these
    three position is shown schematically below.  When the impurity is
    positioned such that the top edge of the kink passes the impurity
    from left to right, the total energy of the system decreases (a).
    The opposite is true if the impurity is on the lower edge of the
    kink (b).}
  \label{fig:impurity-kink}
\end{figure}

Figure \ref{fig:impurity-kink} explains graphically that, depending on
its position, an isolated impurity provides either a left or a right
directed short range force acting on a kink.  In addition to this
force there is a two-kink effect which makes kink pile-ups
energetically favorable when they occur on the impurity site.  

So far we considered only positively charged impurities $\theta = +1$.
The interface is attracted to these impurities.  Negatively charged
impurities with $\theta = -1$ repel the interface.  However, the
qualitative picture of the kink-impurity interaction presented in
Fig.~\ref{fig:impurity-kink} still holds.  The only difference is that
the effect of the negative impurity on the top edge of the kink is
equivalent to the effect of the positive impurity on the bottom edge
and vice versa.  In the limit of high density of impurities,
additional effects due to the interplay of positively and negatively
charged impurities become important.  For example, a row of
alternating positive and negative impurities perpendicular to the
interface pulls the interface along in one direction or another due to
the ease of nucleating kink-antikink pairs.  Additional phenomena
arise when both positively and negatively charged impurities are
presented.  Exploring these these phenomena is outside the scope of
this article.

The diffusing kink picture is especially simple when the impurities
are dilute and $\epsilon \gg J \gg kT$.  In this limit the kinks
diffuse only downslope in the static energy landscape produced by then
impurities.  When impurities are dilute, this energy landscape
consists of a number of flat terraces bound by steep vertical cliffs
or walls.  The kinks diffuse and fall down cliffs until they fall onto
a terrace which is bound by walls on both sides.  The kinks become
trapped on this terrace. In the long time limit, the density of the
trapped kinks becomes uniform on this terrace.  This means that the
slope of the piece of grain boundary which corresponds to this terrace
is a constant given by the density of the trapped kinks.  This density
depends on the initial distribution of kinks and impurities and can be
anything.  Therefore, in this limit, the grain boundary is pinned and
consists of a series of flat facets of random length and inclination.

\begin{figure}[htbp]
  \centering \includegraphics[width=3.3in]{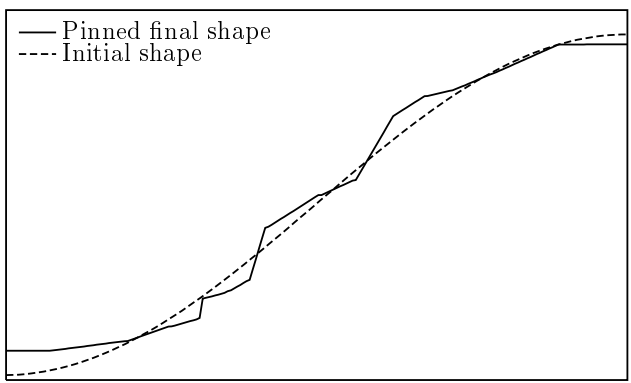}
   \caption{Time averaged shape of a grain boundary pinned by strong
     $\epsilon = 5$ impurities which occupy 1\% of the dual lattice
     sites is a collection of straight segments obtained via a
     Monte-Carlo simulation of the 2d Ising model with an addition
     energy given in Eq.~\eqref{eq:impurities}.}
   \label{fig:piecewise_flat}
\end{figure}

Figure~\ref{fig:piecewise_flat} presents results of the Monte-Carlo
simulation of the low temperature 2d Ising model in the presence of
strong dilute positively charged immobile impurities.  What is shown
is the time-averaged location of the boundary between the spin up and
spin down domains.  The boundary is pinned and consists of straight
pieces of random length and orientation.  Impurities are located at
either end of each such facet.  This result supports our qualitative
picture.

Addition of magnetic field introduces yet another energy scale $H$
into the picture.  When $H \ll \epsilon$, the boundary is pinned.  The
shape of the pinned facets depends on the relative size of $1/\beta H$
and length of the pinned facet $L$.  In equilibrium, the kink drift
due to the magnetic field is balanced by the diffusion due to
curvature.  Thus small facets for which $L \ll 1/\beta H$, remain
straight.  Conversely, when $L \gg 1/\beta H$, the long facets look
like the corners of a droplet expanding in co-aligned magnetic field
whose shape is given in Fig.~\ref{fig:shape}a.

Strong positive impurities ($\epsilon \gg J$) in the bulk of a domain
of aligned spins always have two spins near then that are anti-aligned
with the rest of the spins in that domain.  Thus impurities serve as
nuclei for the formation of droplets of the phase of spins aligned
with the applied magnetic field.  Conversely, strong negatively
charged impurities favor alignment of the nearby spins and thus can
inhibit nucleation of the phase favored by the application of magnetic
field.

\section{Three dimensional Ising model}
\label{sec:3d}

Whereas a curve on a plane can be characterized by a single scalar
curvature, a smooth surface embedded in a three dimensional space is
characterized by a rank two tensor $L_{\alpha\beta}$ ($\{\alpha,
\beta\} = 1, 2$).  This tensor is called the second fundamental form
or the Weingarten map or just the curvature tensor.  The trace of this
tensor is the mean curvature, while its determinant is the Gaussian
curvature.  This tensor is defined at some point $P$ by selecting an
orthogonal coordinate system $x_\alpha$ in the tangent plane at $P$
and writing
\begin{equation}
  \label{eq:L}
  L_{\alpha\beta} = \hat t_\alpha \cdot \frac{\partial\hat n}{\partial
    x_\beta},
\end{equation}
where $\hat t_\alpha$ are the unit tangent vectors, and $\hat n$ is
the unit normal to the surface.

The reduced mobility of a
two-dimensional interface is also a rank two tensor
$M^*_{\alpha\beta}$ which when contracted with the curvature tensor
yields the normal velocity of the interface
\begin{equation}
  \label{eq:v3d}
  v = \sum_{\alpha,\beta} M^*_{\alpha\beta} L_{\beta\alpha}.
\end{equation}
The reduced mobility tensor depends on the scalar bare mobility
$M_\mathrm{3d}$ (found, for example, by measuring the speed of a
driven flat interface) and the interfacial free energy $\gamma$ and
its derivatives.  In the neighborhood of the point $P$ the normal
$\hat n$ is specified by its the deviations $\varphi_1$ and
$\varphi_2$ from the normal at $P$ in the directions $\hat t_1$ and
$\hat t_2$.  The free energy is a function of the normal $\hat n$ and
therefore, in the neighborhood of $P$, a function of these angles
$\gamma_\mathrm{3d}(\varphi_1, \varphi_2)$.  The reduced mobility
tensor is then defined as
\begin{equation}
  \label{eq:Mstar3d}
  M^*_{\alpha\beta} = M_\mathrm{3d}
  \left(
    \gamma \, \delta_{\alpha\beta}  + \frac{\partial^2 \gamma}{\partial
      \varphi_\alpha \partial \varphi_\beta}.
  \right)
\end{equation}

Since our KMC simulations show that the shape of a 3d shrinking grain
is even closer to a sphere than a 2d shape to a circle, this reduced
mobility tensor is nearly isotropic.  This isotropy allows us to
predict the 3d grain shrinking rate (defined as the rate of change of
the $2/3$ power of its volume $S \equiv \frac{d}{dt} \,V^{2/3}$) by
calculating the velocity $v_{100}$ of the shrinking grain boundary at
the $\langle 100 \rangle$ orientation.  We will first estimate this
velocity within the terrace-step-kink description of the vicinal
surface.  We then derive an exact expression for this velocity within
the continuum limit.

\subsection{Shrinking terrace view of the dynamics near $\langle 100
  \rangle$ plane}
\label{sec:3d_terrace_shrink}

\begin{figure}[htbp]
  \centering
  \includegraphics[width=3.3in]{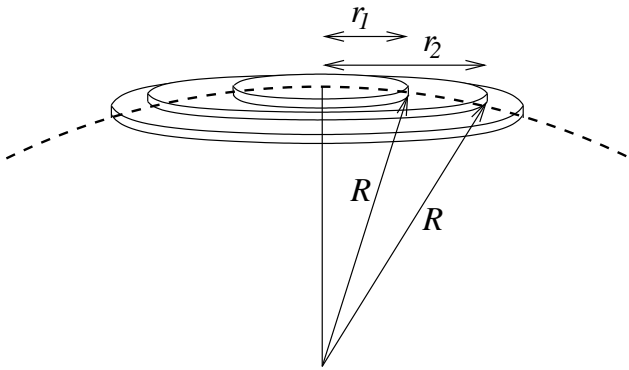}
  \caption{Grain shape near the 100 plane can be viewed as a
    collection of terraces}
  \label{fig:terraces}
\end{figure}
The low temperature interface can be described within the the
terrace-step-kink (TSK) model \cite{burton51:_bcf}.  When steps are
far apart, each step obeys the dynamics of a 2d grain.  If the 3d
grain is a sphere of radius $R$, 
it is described near the $\langle 100 \rangle$
orientation by a stack of circular terraces of
increasing radii $r_1(t)$, $r_2(t)$, etc (See
Fig.~\ref{fig:terraces}).  Because these steps are part of the
spherical grain, they are related via
\begin{equation}
  \label{eq:r1_r2}
  \sqrt{R^2 - r_2^2} + 1 = \sqrt{R^2 - r_1^2}.
\end{equation}
Solving this equation for $R^2$, differentiating with respect to time
and using the exact expression for the 2d grain shrinking rate
\eqref{eq:dAdt} $r_1 \dot r_1 = r_2 \dot r_2 = -2/\pi$ we obtain $R
\dot R = -2/\pi$.  Therefore, within the spherical grain
approximation, the grain shrinking rate is
\begin{equation}
  \label{eq:S_approx}
  S \approx -
  \left(
    \frac{4\pi}{3}
  \right)^{2/3}
  \frac{4}{\pi} \approx -3.309,
\end{equation}
which is in reasonable good quantitative
agreement with the shrinking rate found by KMC
simulations $S = -3.335 \pm 0.001$.

\subsection{Low temperature expansion of interface free energy and mobility
near the $\langle 100 \rangle$ plane}
\label{sec:3d_lowT}

Little analytical progress in deriving equilibrium and kinetic
properties of the 3d KIM has been achieved to date.  A mean field
expression for the free energy of the TSK model neglecting step-step
interaction was obtained by Gruber and Mullins
\cite{gruber67:_anisotropy}.  Holzer and Wortis
\cite{holzer89:_3d_cubic_ising} calculate the free energy near the
$\langle 100 \rangle$ plane in the more controlled diagrammatic
temperature expansion.  It is sufficient for our purposes to keep only
the leading term in the expansion
\begin{equation}
  \label{eq:gamma3d}
  \gamma_\mathrm{3d}(\theta, \phi) \approx \theta \,
  \gamma_\mathrm{2d}(\phi),
\end{equation}
where $\theta$ is the angle between the normal to the interface $\hat
n$ and the $z$-axis (which is the normal to the $\langle 100 \rangle$
plane) and $\phi$ is the angle between the projection of $\hat n$ onto
the $x$-$y$ plane and the $x$-axis counted clockwise.  Formula
\eqref{eq:gamma3d} is simply a statement that the free energy of a
vicinal surface is composed of the free energies of the steps
(considered to be non-interacting).

In the same non-interacting step approximation, the bare mobility of
the vicinal surface is
\begin{equation}
  \label{eq:M3d}
  M_\mathrm{3d}(\theta, \phi) \approx \theta M_\mathrm{2d}(\phi).
\end{equation}

\subsection{Speed of the $\langle 100 \rangle$ orientation of the
  shrinking grain}
\label{sec:3d_M_sharp}

The expressions for the interfacial free energy and the mobility in
the vicinity of the $\langle 100 \rangle$ plane allow us to calculate
the reduced mobility at this orientation as well as the grain shape in
the neighborhood of the point $P$ where the normal is in the
$z$ direction.

Let the shape of the grain in cylindrical coordinates be $z(r, \phi) =
r^2/r^2_\mathrm{2d}(\phi)$.  The shape of the $z = 1$ section of the
3d grain is $r = r_\mathrm{2d}(\phi)$ in polar coordinates.  A
circular terrace, i.e.,\ $r_\mathrm{2d}(\phi) = \sqrt{2R}$,
corresponds to a sphere of radius $R$.  We choose this suggestive
parameterization of the 3d shape with the foresight that
$r_\mathrm{2d}(\phi)$ will turn out to be identical to the shape of
the self-similarly shrinking 2d grain.  This is not surprising in view
of the shrinking terrace picture of the 3d grain evolution near the
$\langle 100 \rangle$ orientation of the previous subsection.

Given the shape of the 3d grain, we can compute the normal of the
surface $\hat n(r, \phi)$ and the curvature tensor $L_{\alpha\beta}(r,
\phi)$.  The reduced mobility tensor $M^*_{\alpha\beta}(\hat n)$, in
turn, can be calculated using the free energy expression
\eqref{eq:gamma3d} following the prescription \eqref{eq:Mstar3d}.
Their contraction is the local velocity of the interface $v(r, \phi)$.
We will not write down the full expression for $v$ due to its
unenlightening complexity.  It's $r \rightarrow 0$ limit has to be
independent of the direction of approach $\phi$.  We obtain
\begin{multline}
  \label{eq:v3d_phi}
  \lim_{r\rightarrow 0} v(r, \phi) = \cr -\frac{2 \,
  [r_\mathrm{2d}^2(\phi) + 2 r'_\mathrm{2d}(\phi) -
  r''_\mathrm{2d}(\phi)]}{r_\mathrm{2d}^2(\phi)
  \left[
    r_\mathrm{2d}(\phi) (\sin\phi + \cos\phi)
    + r'_\mathrm{2d}(\phi) (\sin\phi - \cos\phi)
  \right]^2} \cr = const,
\end{multline}
subject to the smoothness constraints $r'_\mathrm{2d}(0) =
r'_\mathrm{2d}(\phi) = 0$.  The function $r_\mathrm{2d}(\phi)$ which
satisfies the above equation and constraints is precisely the
self-similar shape of a 2d shrinking grain.  If we set the size of the
3d grain by choosing $r_\mathrm{2d}(\pi/4) = \sqrt{2R}$, we arrive at
$v = -2B/R$ and hence (again assuming a spherical shape to estimate
the volume)
\begin{equation}
  \label{eq:S_approx1}
  S \approx -
  \left(
    \frac{4\pi}{3}
  \right)^{2/3} 4B \approx -3.445.
\end{equation}
Since the diameter of the self-similar shape at $\phi = 0$ is slightly
smaller than its diameter at $\phi = \pi/4$, a better approximation
would have been to scale $r_\mathrm{2d}$ in such a way that at the
intermediate angle $r_\mathrm{2d}(\pi/8) = \sqrt{2R}$.

\section{Conclusions}
\label{sec:conclusions}
In summary, using a kink-based description, we have derived directly
from a microscopic model (low temperature KIM) a continuum evolution
equation for the anisotropic motion of a simple interface, and we have
shown its equivalence to the standard phenomenological law of motion
by curvature.  We have illustrated with the example of dilute
impurities that this kink-based kinetic description provides a useful
framework for studying more complex situations.  By extending this
description to 3d, and by exploiting our 2d result for the
self-similar dynamics of shrinking terraces, we have obtained the
velocity of a curved interface near a singular orientation. We have
shown that even though the interface stiffness tensor and the
curvature tensor are singular at the $\langle 100\rangle$ orientation,
their product, which determines the interface velocity, is smooth.
Furthermore, this velocity is consistent with the one predicted from
the 3d tensorial generalization of the law for anisotropic
curvature-driven motion using known expressions for the interface free
energy and bare mobility.

Our kink-based derivation of a continuum equation of interface motion
highlights the microscopic mechanism for the remarkable isotropy of
the reduced mobility in both 2d and 3d and thus the shape of grains
shrinking under the influence of capillarity alone.  The reduced
mobility is a product of the interfacial stiffness and the interfacial
mobility both of which are strongly anisotropic.  The isotropy of the
reduced mobility is therefore a result of the cancellation of
anisotropies of the interfacial stiffness and interfacial mobility.
The microscopic reason for the cancellation is purely geometric in
origin.  The number of geometrically necessary kinks, and hence the
configurational entropy of the interface, varies rapidly with
inclination near low-energy/low-mobility orientations, but slowly near
high-energy/high-mobility interfaces, where the density of kinks is
high.  Since the leading order contribution to the interfacial
stiffness comes from configurational entropy, stiffness is high where
mobility is low and vice versa.  The cancellation of anisotropies
leads to roughly isotropic reduced mobility.  Therefore the shape of a
shrinking grain can appear isotropic or anisotropic depending on
whether driving forces other than capillarity are present.  The bare
mobility of the interface is, however, independent of the nature of
the driving force.

An interesting prospect for the future is to extend this kink-based
theoretical description of interface motion to realistic, and more
complex, grain boundaries where kinks have the character of
dislocations.  Work along this line is presently in progress.

We thank Bernard Derrida, Mikhail Mendelev, Anthony Rollet, and David
Srolovitz for valuable discussions.  This research is supported by
U.S.\ DOE through Grant No.~DE-FG02-92ER45471 and funds from the
Computational Materials Science Network.

\appendix

\section{Derivation of the kink equation of motion}
\label{app:derivation}

In general, kinks comprising the grain boundary are characterized by
their width $b$, which is the distance of the closest approach of two
neighboring kinks, their height $d$, and the length of the steps of
their random walk $a$.  For example, on square lattice, $d = a$, the
lattice constant and $b = 0$, while on a triangular lattice $d =
a\sqrt{3}/2$ and $b = a/2$.  In the continuum limit, we define the
density of kinks $\rho(x, t)$ and seek its evolution equation in some
fixed domain $x \in [x_\textrm{L}, x_\mathrm{R}]$.  Since neighboring
domains contain anti-kinks, absorbing boundary conditions must be
imposed $\rho(x_\mathrm{L}, t) = \rho(x_\mathrm{R}, t) = 0$.

A special case of this problem $a = b$ describes random walkers in 1d
which cannot occupy the same site.  When a magnetic field is present,
the random walk is biased and the problem can be mapped
onto the well studied asymmetric exclusion
process \cite{liggett85:_interacting,derrida92:_exact}.  To map
the problem of finding the
evolution of the kink density $\rho(x,t)$ onto this special
problem, we insert a space $c = a - b$ between each pair of adjacent
kinks, as illustrated for a square lattice in Fig.~\ref{fig:mapping}.
The resulting kink density $R(\xi, t) \in [0, 1/a]$ is defined in a
different domain $\xi \in [\xi_\mathrm{L}(t), \xi_\mathrm{R}(t)]$.
In the presence of magnetic field $H$, this
kink density satisfies the equation
\cite{liggett85:_interacting,spohn91:_large_scale} (subscripts denote
differentiation)
\begin{equation}
  \label{eq:Rdot}
  R_t = D R_{\xi\xi} + \alpha [R(1 - aR)]_\xi = -J_\xi,
\end{equation}
with
\begin{equation}
  \label{eq:J}
  J(\xi, t) = -D R_\xi - \alpha R(1 - aR),
\end{equation}
where $D = a^2/2\tau$, $\alpha = a\beta H/\tau$, $\tau$ is the Monte
Carlo time step, and $J$ is the flux of kinks in the moving
$\xi$-domain. Note that this equation is identical to Burger's equation
after elimination of the drift term $\alpha R_\xi$ by transformation to a
moving frame.  As kinks annihilate at the boundaries of the
$x$-domain, the $\xi$-domain shrinks.  Each kink that leaves the
$x$-domain, decreases the $\xi$-domain by $c$.  This implies that the
boundaries of the $\xi$-domain move with a velocities proportional to
the current of kinks our of the domain
\begin{equation}
  \label{eq:xi_dot}
  \dot \xi_\mathrm{L} = -c \, J(\xi_\mathrm{L}(t), t), \quad
  \dot \xi_\mathrm{R} = -c \, J(\xi_\mathrm{R}(t), t).
\end{equation}
The equations for the motion of boundaries \eqref{eq:xi_dot}, together
with the absorbing boundary conditions
\begin{equation}
  \label{eq:R_absorbing}
  R(\xi_\mathrm{L}(t), t) = R(\xi_\mathrm{R}(t), t) = 0,
\end{equation}
completely define the problem of diffusing kinks in the $\xi$-domain.

The mapping is inverted via
\begin{equation}
  \label{eq:xi_x}
  \xi(x, t) - \xi_\mathrm{L}(t) = x - x_\mathrm{L}+
  c\int_{x_\mathrm{L}}^x \rho(x', t) \, dx'.
\end{equation}
At some fixed time we can write
\begin{equation}
  \label{eq:physical_point}
  R(\xi, t) d\xi = \rho(x, t) \, dx,
\end{equation}
since both expressions give the number of kinks in the same physical
interval.  Using \eqref{eq:xi_x} we obtain
\begin{equation}
  \label{eq:R_rho}
  R(\xi, t) = \frac{\rho(x, t)}{1 + c \rho(x, t)}, \quad \mathrm{or}
  \quad \rho(x, t) = \frac{R(\xi, t)}{1 - c R(\xi, t)}.
\end{equation}
This relationship \eqref{eq:physical_point} allows us to invert
\eqref{eq:xi_x} to obtain
\begin{equation}
  \label{eq:x_xi}
  x(\xi, t) - x_\mathrm{L} = \xi - \xi_\mathrm{L}(t) -
  c\int_{\xi_\mathrm{L}(t)}^\xi R(\xi', t) \, d\xi'.
\end{equation}
It is now only a matter of carrying out the chain rule together with
the boundary conditions \eqref{eq:R_absorbing} and the transformation
\eqref{eq:xi_x} to obtain
\begin{subequations}
  \label{eq:R_derivs}
  \begin{eqnarray}
    \label{eq:R_xi}
    R_\xi &=& \frac{\rho_x}{(1 + c\rho)^3}, \\
    \label{eq:R_t}
    R_t &=& \frac{\rho_t}{(1 + c\rho)^2} - D \, \frac{c\rho_x^2}{(1 +
      c\rho)^5} \nonumber \\ && -\alpha \, \frac{\rho_x \, c\rho(1 -
      b\rho)}{(1 + c\rho)^4}, \\
    \label{eq:R_xixi}
    R_{\xi\xi} &=& \frac{\rho_{xx}}{(1 + c\rho)^4} -
      \frac{3c\rho_x^2}{(1 + c\rho)^5}.    
  \end{eqnarray}
\end{subequations}
Thus we obtain the non-linear diffusion equation for $\rho(x, t)$
which reads
\begin{multline}
  \label{eq:rho_t_general}
  \rho_t = D \, \frac{\rho_{xx}}{(1 + c\rho)^2} - 2cD\,
    \frac{\rho_x^2}{(1 + c\rho)^3} + \cr \alpha \, \frac{\rho_x(1 -
    2b\rho - bc \rho^2)}{(1 + c\rho)^2} = - F_x,
\end{multline}
where 
\begin{equation}
  \label{eq:kink_flux}
  F = -D \,\frac{\rho_x}{(1 + c\rho)^2} - \alpha \rho \, 
    \frac{1 - b\rho}{1 + c\rho},
\end{equation}
is the flux of kinks in the fixed domain which vanishes at zero kink
density.

Using the relationship of the local slope and kink density $\tan \phi
= h_x(x, t) = d\, \rho(x, t)$ and the expressions for the normal
interface velocity and curvature
\begin{equation}
  \label{eq:v_n_curvature}
    v_\mathrm{n} = \frac{h_t}{(1 + h_x^2)^{1/2}}, \quad   \kappa =
    \frac{h_{xx}}{(1 + h_x^2)^{3/2}},
\end{equation}
we can compute the bare and the reduced mobilities from the normal
velocity of the interface $v_n = M^* \kappa + M H$.  We obtain
\begin{equation}
  \label{eq:Mstar_general}
  M^* = \frac{D}{(\cos\phi + \nu\sin\phi)^2}, \quad M =
  \frac{\beta}{\tau} \, \frac{\lambda \sin\phi \, (\cos\phi -
  \mu\sin\phi)}{\cos\phi + \nu\sin\phi},
\end{equation}
where $\nu = c/d$, $\mu = b/d$ and $\lambda = a/d$ are geometric
factors.  These expressions are valid for $\phi \in [0,\pi/4]$ for the
square lattice and for $\phi \in [0, \pi/6]$ for the triangular
lattice.

\section{Self-similar shrinking grain on a hexagonal lattice}
\label{sec:self-similar_hex}

The symmetry of the hexagonal lattice allows us to solve for the shape
of the self-similarly shrinking grain in a $\pi/6$ wedge.  We present
yet another way of obtaining this shape.  Let the points on the
boundary be labeled by $\phi$, the azimuthal angle $\phi \in [0,
\pi/6]$.  Let $\theta(\phi)$ be the local slope and
$r_\mathrm{2d}(\phi)$ the radial distance from the center of the
grain.  The shrinking shape will remain self-similar if the radial
velocity $v_r$ at each point of the boundary is proportional to the
radius at that point.  The normal velocity $v_n = M^* \kappa$ is the
projection of the radial boundary velocity onto the normal direction.
The curvature is the derivative of the slope with respect to the arc
length
\begin{equation}
  \label{eq:curvature_arc_length}
  \kappa = \frac{d\theta}{ds} = \frac{\theta'\cos(\phi -
    \theta)}{r_\mathrm{2d}}.
\end{equation}
Thus, the condition of the self-similarity of the shrinking shape can
be written as
\begin{equation}
  \label{eq:radial_vel}
  v_r = \frac{v_n}{\cos(\phi - \theta)} = M^* \,
  \frac{\theta'}{r_\mathrm{2d}} = C \, r_\mathrm{2d},
\end{equation}
where $C$ is some proportionality constant.  To complete the
description we need to express the radius $r_\mathrm{2d}(\phi)$ in
terms of $\theta(\phi)$
\begin{equation}
  \label{eq:r_2d_theta}
  r_\mathrm{2d}'(\phi) = r_\mathrm{2d}(\phi) \, \sin(\phi - \theta(\phi)).
\end{equation}
Without loss of generality we set $r_\mathrm{2d}(0) = 1$ and integrate
equations \eqref{eq:radial_vel} and \eqref{eq:r_2d_theta} together up
to $\phi = \pi/6$.  The second boundary $\theta$-boundary condition
$\theta(\pi/6) = \pi/6$ selects a unique $C$.  The numerical shooting
yields $C \approx 0.903535$.  The shape of the self-similarly shrinking
grain on a hexagonal lattice is remarkably close to a circle.  The
largest and smallest grain diameters differ by only 0.4\%!


\begin{thebibliography}{24}
\expandafter\ifx\csname natexlab\endcsname\relax\def\natexlab#1{#1}\fi
\expandafter\ifx\csname bibnamefont\endcsname\relax
  \def\bibnamefont#1{#1}\fi
\expandafter\ifx\csname bibfnamefont\endcsname\relax
  \def\bibfnamefont#1{#1}\fi
\expandafter\ifx\csname citenamefont\endcsname\relax
  \def\citenamefont#1{#1}\fi
\expandafter\ifx\csname url\endcsname\relax
  \def\url#1{\texttt{#1}}\fi
\expandafter\ifx\csname urlprefix\endcsname\relax\def\urlprefix{URL }\fi
\providecommand{\bibinfo}[2]{#2}
\providecommand{\eprint}[2][]{\url{#2}}

\bibitem[{\citenamefont{Winning et~al.}(2002)\citenamefont{Winning, Gottstein,
  and Shvindlerman}}]{winning02:_mechanisms}
\bibinfo{author}{\bibfnamefont{M.}~\bibnamefont{Winning}},
  \bibinfo{author}{\bibfnamefont{G.}~\bibnamefont{Gottstein}},
  \bibnamefont{and} \bibinfo{author}{\bibfnamefont{L.~S.}
  \bibnamefont{Shvindlerman}}, \bibinfo{journal}{Acta Mat.}
  \textbf{\bibinfo{volume}{50}}, \bibinfo{pages}{353} (\bibinfo{year}{2002}).

\bibitem[{\citenamefont{Molodov et~al.}(1999)\citenamefont{Molodov, Gottstein,
  Heringhaus, and Shvindlerman}}]{molodov99:_magnet}
\bibinfo{author}{\bibfnamefont{D.~A.} \bibnamefont{Molodov}},
  \bibinfo{author}{\bibfnamefont{G.}~\bibnamefont{Gottstein}},
  \bibinfo{author}{\bibfnamefont{F.}~\bibnamefont{Heringhaus}},
  \bibnamefont{and} \bibinfo{author}{\bibfnamefont{L.~S.}
  \bibnamefont{Shvindlerman}}, \bibinfo{journal}{Mat. Sci. Forum}
  \textbf{\bibinfo{volume}{294}}, \bibinfo{pages}{127} (\bibinfo{year}{1999}).

\bibitem[{\citenamefont{Herring}(1949)}]{herring49}
\bibinfo{author}{\bibfnamefont{C.}~\bibnamefont{Herring}}, in
  \emph{\bibinfo{booktitle}{The Physics of Powder Metallurgy}}, edited by
  \bibinfo{editor}{\bibfnamefont{W.~E.} \bibnamefont{Kingston}}
  (\bibinfo{publisher}{McGraw Hill}, \bibinfo{address}{New York, NY},
  \bibinfo{year}{1949}).

\bibitem[{\citenamefont{Hillert}(1983)}]{hillert83:_digm}
\bibinfo{author}{\bibfnamefont{M.}~\bibnamefont{Hillert}},
  \bibinfo{journal}{Scripta Metal.} \textbf{\bibinfo{volume}{17}},
  \bibinfo{pages}{237} (\bibinfo{year}{1983}).

\bibitem[{\citenamefont{Huang and Humphreys}(2001)}]{huang01:_solute}
\bibinfo{author}{\bibfnamefont{Y.}~\bibnamefont{Huang}} \bibnamefont{and}
  \bibinfo{author}{\bibfnamefont{F.~J.} \bibnamefont{Humphreys}}, in
  \emph{\bibinfo{booktitle}{Proc. 1$^\mathrm{st}$ Intl. Conf. on
  Recrystallization and Grain Growth}} (\bibinfo{publisher}{Springer},
  \bibinfo{address}{Aachen, Germany}, \bibinfo{year}{2001}), pp.
  \bibinfo{pages}{409--14}.

\bibitem[{\citenamefont{Molodov et~al.}(1998)\citenamefont{Molodov, Gottstein,
  Heringhaus, and Shvindlerman}}]{molodov98:_true}
\bibinfo{author}{\bibfnamefont{D.~A.} \bibnamefont{Molodov}},
  \bibinfo{author}{\bibfnamefont{G.}~\bibnamefont{Gottstein}},
  \bibinfo{author}{\bibfnamefont{F.}~\bibnamefont{Heringhaus}},
  \bibnamefont{and} \bibinfo{author}{\bibfnamefont{L.~S.}
  \bibnamefont{Shvindlerman}}, \bibinfo{journal}{Acta Mat.}
  \textbf{\bibinfo{volume}{46}}, \bibinfo{pages}{5627} (\bibinfo{year}{1998}).

\bibitem[{\citenamefont{Mendelev et~al.}(2002)\citenamefont{Mendelev,
  Srolovitz, Gottstein, and Shvindlerman}}]{mendelev02:_driving}
\bibinfo{author}{\bibfnamefont{M.~I.} \bibnamefont{Mendelev}},
  \bibinfo{author}{\bibfnamefont{D.~J.} \bibnamefont{Srolovitz}},
  \bibinfo{author}{\bibfnamefont{G.}~\bibnamefont{Gottstein}},
  \bibnamefont{and} \bibinfo{author}{\bibfnamefont{L.~S.}
  \bibnamefont{Shvindlerman}}, \bibinfo{journal}{J. Mater. Res.}
  \textbf{\bibinfo{volume}{17}}, \bibinfo{pages}{234} (\bibinfo{year}{2002}).

\bibitem[{\citenamefont{Lobkovsky et~al.}(2004)\citenamefont{Lobkovsky, Karma,
  Mendelev, Haataja, and Srolovitz}}]{lobkovsky04:_grain_shape}
\bibinfo{author}{\bibfnamefont{A.~E.} \bibnamefont{Lobkovsky}},
  \bibinfo{author}{\bibfnamefont{A.}~\bibnamefont{Karma}},
  \bibinfo{author}{\bibfnamefont{M.~I.} \bibnamefont{Mendelev}},
  \bibinfo{author}{\bibfnamefont{M.}~\bibnamefont{Haataja}}, \bibnamefont{and}
  \bibinfo{author}{\bibfnamefont{D.~J.} \bibnamefont{Srolovitz}},
  \bibinfo{journal}{Acta. Mat.} \textbf{\bibinfo{volume}{52}},
  \bibinfo{pages}{285} (\bibinfo{year}{2004}).

\bibitem[{\citenamefont{Glauber}(1963)}]{glauber63:ising}
\bibinfo{author}{\bibfnamefont{R.~J.} \bibnamefont{Glauber}},
  \bibinfo{journal}{J. Math. Phys.} \textbf{\bibinfo{volume}{4}},
  \bibinfo{pages}{294} (\bibinfo{year}{1963}).

\bibitem[{\citenamefont{Rottman and Wortis}(1981)}]{rottman81:_exact}
\bibinfo{author}{\bibfnamefont{C.}~\bibnamefont{Rottman}} \bibnamefont{and}
  \bibinfo{author}{\bibfnamefont{M.}~\bibnamefont{Wortis}},
  \bibinfo{journal}{Phys. Rev. B} \textbf{\bibinfo{volume}{24}},
  \bibinfo{pages}{6274} (\bibinfo{year}{1981}).

\bibitem[{\citenamefont{Avron et~al.}(1982)\citenamefont{Avron, Vanbeijeren,
  Schulman, and Zia}}]{avron82:_roughening}
\bibinfo{author}{\bibfnamefont{J.~E.} \bibnamefont{Avron}},
  \bibinfo{author}{\bibfnamefont{H.}~\bibnamefont{Vanbeijeren}},
  \bibinfo{author}{\bibfnamefont{L.~S.} \bibnamefont{Schulman}},
  \bibnamefont{and} \bibinfo{author}{\bibfnamefont{R.~K.~P.}
  \bibnamefont{Zia}}, \bibinfo{journal}{J. Phys. A}
  \textbf{\bibinfo{volume}{15}}, \bibinfo{pages}{L81} (\bibinfo{year}{1982}).

\bibitem[{\citenamefont{Kochmanski}(1999)}]{kochmanski99:_ising_asympt}
\bibinfo{author}{\bibfnamefont{M.~S.} \bibnamefont{Kochmanski}},
  \bibinfo{journal}{J. Phys. A: Math. Gen.} \textbf{\bibinfo{volume}{32}},
  \bibinfo{pages}{1251} (\bibinfo{year}{1999}).

\bibitem[{\citenamefont{Yurishchev}(1997)}]{yurishchev97:_critic_ising}
\bibinfo{author}{\bibfnamefont{M.~A.} \bibnamefont{Yurishchev}},
  \bibinfo{journal}{Phys. Rev. E} \textbf{\bibinfo{volume}{55}},
  \bibinfo{pages}{3915} (\bibinfo{year}{1997}).

\bibitem[{\citenamefont{Spohn}(1993)}]{spohn93:_interface}
\bibinfo{author}{\bibfnamefont{H.}~\bibnamefont{Spohn}}, \bibinfo{journal}{J.
  Stat. Phys.} \textbf{\bibinfo{volume}{71}}, \bibinfo{pages}{1081}
  (\bibinfo{year}{1993}).

\bibitem[{\citenamefont{Barma}(1992)}]{barma92:_dynam_ising}
\bibinfo{author}{\bibfnamefont{M.}~\bibnamefont{Barma}}, \bibinfo{journal}{J.
  Phys. A} pp. \bibinfo{pages}{L693--9} (\bibinfo{year}{1992}).

\bibitem[{\citenamefont{Rikvold and Kolesik}(2000)}]{rikvold00:_analytic}
\bibinfo{author}{\bibfnamefont{P.~A.} \bibnamefont{Rikvold}} \bibnamefont{and}
  \bibinfo{author}{\bibfnamefont{M.}~\bibnamefont{Kolesik}},
  \bibinfo{journal}{J. Stat. Phys.} \textbf{\bibinfo{volume}{100}},
  \bibinfo{pages}{377} (\bibinfo{year}{2000}).

\bibitem[{\citenamefont{Liggett}(1985)}]{liggett85:_interacting}
\bibinfo{author}{\bibfnamefont{T.~M.} \bibnamefont{Liggett}},
  \emph{\bibinfo{title}{Interacting particle systems}}
  (\bibinfo{publisher}{Springer-Verlag}, \bibinfo{address}{New York},
  \bibinfo{year}{1985}).

\bibitem[{\citenamefont{Masi et~al.}(1984)\citenamefont{Masi, Ianiro,
  Pellegrinotti, and Presutti}}]{masi84:_nonequilibrium}
\bibinfo{author}{\bibfnamefont{A.~D.} \bibnamefont{Masi}},
  \bibinfo{author}{\bibfnamefont{N.}~\bibnamefont{Ianiro}},
  \bibinfo{author}{\bibfnamefont{A.}~\bibnamefont{Pellegrinotti}},
  \bibnamefont{and} \bibinfo{author}{\bibfnamefont{E.}~\bibnamefont{Presutti}},
  in \emph{\bibinfo{booktitle}{Nonequilibrium Phenomena II: from Stochastics to
  Hydrodynamics}}, edited by \bibinfo{editor}{\bibfnamefont{J.~L.}
  \bibnamefont{Lebowitz}} \bibnamefont{and}
  \bibinfo{editor}{\bibfnamefont{E.~W.} \bibnamefont{Montroll}}
  (\bibinfo{publisher}{Elsevier}, \bibinfo{address}{New York},
  \bibinfo{year}{1984}).

\bibitem[{\citenamefont{Kandel and Domany}(1990)}]{kandel90:_rigorous}
\bibinfo{author}{\bibfnamefont{D.}~\bibnamefont{Kandel}} \bibnamefont{and}
  \bibinfo{author}{\bibfnamefont{E.}~\bibnamefont{Domany}},
  \bibinfo{journal}{J. Stat. Phys.} \textbf{\bibinfo{volume}{58}},
  \bibinfo{pages}{685} (\bibinfo{year}{1990}).

\bibitem[{\citenamefont{Burton et~al.}(1951)\citenamefont{Burton, Cabrera, and
  Frank}}]{burton51:_bcf}
\bibinfo{author}{\bibfnamefont{W.~K.} \bibnamefont{Burton}},
  \bibinfo{author}{\bibfnamefont{N.}~\bibnamefont{Cabrera}}, \bibnamefont{and}
  \bibinfo{author}{\bibfnamefont{F.~C.} \bibnamefont{Frank}},
  \bibinfo{journal}{Phil. Trans. R. Soc. London, Ser. A}
  \textbf{\bibinfo{volume}{243}}, \bibinfo{pages}{299} (\bibinfo{year}{1951}).

\bibitem[{\citenamefont{Gruber and Mullins}(1967)}]{gruber67:_anisotropy}
\bibinfo{author}{\bibfnamefont{E.~E.} \bibnamefont{Gruber}} \bibnamefont{and}
  \bibinfo{author}{\bibfnamefont{W.~W.} \bibnamefont{Mullins}},
  \bibinfo{journal}{J. Phys. Chem. Solids} \textbf{\bibinfo{volume}{28}},
  \bibinfo{pages}{875} (\bibinfo{year}{1967}).

\bibitem[{\citenamefont{Holzer and Wortis}(1989)}]{holzer89:_3d_cubic_ising}
\bibinfo{author}{\bibfnamefont{M.}~\bibnamefont{Holzer}} \bibnamefont{and}
  \bibinfo{author}{\bibfnamefont{M.}~\bibnamefont{Wortis}},
  \bibinfo{journal}{Phys. Rev. B} \textbf{\bibinfo{volume}{40}},
  \bibinfo{pages}{11044} (\bibinfo{year}{1989}).

\bibitem[{\citenamefont{Derrida et~al.}(1992)\citenamefont{Derrida, Domany, and
  Mukamel}}]{derrida92:_exact}
\bibinfo{author}{\bibfnamefont{B.}~\bibnamefont{Derrida}},
  \bibinfo{author}{\bibfnamefont{E.}~\bibnamefont{Domany}}, \bibnamefont{and}
  \bibinfo{author}{\bibfnamefont{D.}~\bibnamefont{Mukamel}},
  \bibinfo{journal}{J. Stat. Phys.} \textbf{\bibinfo{volume}{69}},
  \bibinfo{pages}{667} (\bibinfo{year}{1992}).

\bibitem[{\citenamefont{Spohn}(1991)}]{spohn91:_large_scale}
\bibinfo{author}{\bibfnamefont{H.}~\bibnamefont{Spohn}},
  \emph{\bibinfo{title}{Large Scale Dynamics of Interacting Particles}}
  (\bibinfo{publisher}{Springer-Verlag}, \bibinfo{address}{New York},
  \bibinfo{year}{1991}).

\end{thebibliography}
\end{document}